\documentclass[Physsubmission, Phys]{SciPost}

\hypersetup{
    colorlinks,
    linkcolor={red!50!black},
    citecolor={blue!50!black},
    urlcolor={blue!80!black}
}

\usepackage[bitstream-charter]{mathdesign}
\DeclareSymbolFont{usualmathcal}{OMS}{cmsy}{m}{n}
\DeclareSymbolFontAlphabet{\mathcal}{usualmathcal}

\begin{document}

\begin{center}{\Large \textbf{
Application of Lorentzian CFT Principal Series Representation to Near Forward Scattering
}}\end{center}

\begin{center}
Pulkit Agarwal\textsuperscript{1},
Richard C. Brower\textsuperscript{2},
Timothy G. Raben\textsuperscript{3} and
Chung-I Tan\textsuperscript{4}
\end{center}

\begin{center}
{\bf 1} National University of Singapore, Singapore
\\
{\bf 2} Boston University, Boston, MA 02215
\\
{\bf 3} Michigan State University
\\
{\bf 4} Brown University, Providence, RI 02912
\\

\end{center}

\begin{center}
\today
\end{center}


\definecolor{palegray}{gray}{0.95}
\begin{center}
\colorbox{palegray}{
  \begin{tabular}{rr}
  \begin{minipage}{0.1\textwidth}
    \includegraphics[width=30mm]{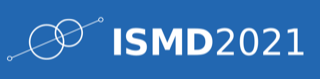}
  \end{minipage}
  &
  \begin{minipage}{0.75\textwidth}
    \begin{center}
    {\it 50th International Symposium on Multiparticle Dynamics}\\ {\it (ISMD2021)}\\
    {\it 12-16 July 2021} \\
    \doi{10.21468/SciPostPhysProc.?}\\
    \end{center}
  \end{minipage}
\end{tabular}
}
\end{center}

\section*{Abstract}
{\bf
We present a discussion on recent progress in high energy diffraction from the perspective of AdS/CFT, through which a unified treatment for both perturbative and non-perturbative Pomeron emerges. By working with Unitary Irreducible Representation of Conformal group, a frame is provided in extending AdS/CFT to both forward and near-forward scattering. We present an analysis involving an exact solution to conformal blocks in Minkowski CFT and discuss possible applications. Phenomenological applications can range from forward scattering to DIS/DVCS/TMD at LHC energies and beyond.
}

\pagenumbering{gobble}
\pagenumbering{arabic}
\section{Introduction}
\label{sec:intro}
Study of Conformal Field Theories (CFTs) has recently gained traction within various domains of physics. Quantum Chromodynamics (QCD) has numerous limits where the dynamics appear near conformal \cite{Braun:2003rp}. The use of conformal methods for QCD are numerous\footnote{We highlight here a very brief selection of a vast literature.} and prove especially useful in the high energy limit away from any relevant mass scale (e.g. \cite{lipatov1993high}),and in the strong coupling limit using holography (e.g. \cite{Maldacena:1998im,Witten:1998qj}). The latter example is of particular interest where a unified description of the perturbative and non-perturbative Pomeron emerges \cite{Brower:2006ea}. The canonical AdS/CFT approach is formulated using Euclidean CFTs. In order to make claims about systems like particle scattering, the results would have to be analytically continued to Lorentzian space---and many nontrivial connections between Euclidean and Minkowski CFTs have been found \cite{Caron-Huot:2017vep,Kravchuk:2018htv}. A direct treatment of Minkowski CFTs was done in \cite{Raben:2018rbn}, which emphasized the importance of Regge behavior for the associated conformal blocks. It deals with the study of scattering within the CFT framework where it was shown that an invariant function over $ SO(d,2) $ can be represented via the unitary principal series as
\begin{equation}\label{eq:unitarySA}
	F(g) = \sum_{m,m'}  \int_{-i \infty}^{i \infty} \frac {d \widetilde \ell }{2\pi i} \int_{-i \infty}^{i \infty} \frac {d \widetilde  \Delta }{2\pi i} \, \, a_{m,m'}(\widetilde  \ell ,\widetilde  \Delta) \,{\mathcal G}_{(\widetilde\ell,\widetilde \Delta; m,m')} (g) .
\end{equation}
The representation functions (or group harmonics) $ \mathcal{G}_{\Delta,\ell}(g) $ can be associated with the defined Minkowski conformal blocks. 

This shift from a Euclidean metric to a Minkowski one can be thought of as moving to a different real form of the same complex extension: $ SO(d+1,1)\rightarrow SO(d,2) $. The aim is to provide an unitary, irreducible representation for $SO(d,2)$ using a convenient induced representation. Under Iwasawa decomposition, $ G=KAN $, \cite{knapp2002lie}, there is a rank 2 Maximum Abelian Subgroup $ A = SO(1,1)\times SO(1,1)$ of $ SO(d,2) $ which carries the physics of the two independent non-compact generators in the group. These generators are chosen to be a dilatation and a longitudinal Lorentz boost. For CFT, the group harmonics $ \mathcal{G}_{\Delta,\ell}(g) $, which are purely kinematical, are found through standard procedure via the unitarized Casimir equation using appropriate Regge boundary conditions and are a function of the two non-compact parameters (which will be related to the two familiar conformal cross ratios). 
These group harmonics are generalized spherical functions that are obtained after integrating over the maximal compact subgroup $K=SO(d)\times SO(2)$, thus reducing Eq. (\ref{eq:unitarySA}) to the limit of $m=m'=0$ only \cite{Agarwal:2021}.

With applications to diffractive physics as our goal, we will start by setting up the relevant kinematics. We will discuss how the two conformal cross ratios can be combined to make connection with our group theoretic interpretation. After "diagonalizing" the Casimir operator, we will state our solutions of the group harmonics for the special cases of $ d=2,4 $, which will be most relevant for phenomenological applications to diffractive physics. We will conclude with a brief discussion of these applications.

\section{Kinematics}
In order to discuss scattering in CFT, the process we consider is that of off-shell scattering of four-currents\footnote{For illustrative purpose, we shall treat external states as conformal scalars.}
\begin{equation}\label{eq:scatter}
	\gamma^\star(1) + \gamma^\star(3) \rightarrow \gamma^\star(2) + \gamma^\star(4).
\end{equation}
Scattering amplitudes for such a process can be written in terms of a partial wave decomposition where the dynamics are encoded in the partial wave amplitudes $a(\ell,\Delta)$. These amplitudes can in principle be found exactly in a theory with a Lagrangian. In a theory without a Lagrangian, like our Regge treatment, a fundamental physical assumption is that of meromorphy. The success of a CFT treatment of QCD lies in an agreement between these two approaches.

In this set-up, we can choose a causal structure that makes studying the $t$-channel OPE most natural as follows. For the $ s $-channel scattering process Eq. \ref{eq:scatter}, points $ (1,4) $ and $ (2,3) $ are timelike separated\footnote{In contrast, all distances in a Euclidean setting are necessarily spacelike separated.} $(x^2_{14},x^2_{23}<0 )$ whereas all other distances are spacelike (eg. $x_{13}^2, x_{24}^2 >0$). Therefore, we designate our $ 2\rightarrow2 $ scattering $ s $-channel process as $ 1+3\rightarrow 2+4 $. The corresponding process $1+\bar 2\rightarrow \bar 3+4$ is the $ t $-channel and $1+\bar 4\rightarrow 2+\bar 3$ is $ u $-channel. In terms of cross ratios: $u=({x_{12}^2} {x_{34}^2})/( {x_{13}^2}{x_{24}^2})$ and  $v= ({x_{23}^2}{x_{14}^2})/({x_{13}^2}{x_{24}^2})$ we are restricted to the first quadrant $(u,v>0)$. Since this is also the case for Euclidean CFT, it is useful to consider instead the $ (\sqrt{u},\sqrt{v}) $\footnote{The $(u,v)$ plane has a double sheet structure that the $(\sqrt{u},\sqrt{v})$ plane resolves. } plane such that $\sqrt u=(\sqrt {x_{12}^2} \sqrt{x_{34}^2})/(\sqrt {x_{13}^2}\sqrt{x_{24}^2})$ and $\sqrt v=(\sqrt{x_{23}^2}\sqrt{x_{14}^2})/(\sqrt{x_{13}^2}\sqrt{x_{24}^2})$.

Notice that we now cover a larger part of the $ (\sqrt{u},\sqrt{v}) $ plane since, for the $ s $-channel say, $ \sqrt{v}<0 $\footnote{We have chosen here a "positive root" convention meaning that $ \sqrt{x^2_{ij}}>0 $. Similar procedure can be followed for the $ u $-channel to find $ \sqrt{u},\sqrt{v}<0 $.}. We can define a set of crossing symmetric variables $ w,\sigma $ by requiring that $ w\sigma = \frac{1-v}{u}$. For our purposes we choose\footnote{A similar choice was made in the Euclidean lattice treatment of \cite{Brower:2020jqj}.}: $w\equiv \frac{1-\sqrt v}{\sqrt u}$ and $\sigma \equiv  \frac{1+\sqrt v}{\sqrt u}$. In order to see why this set is interesting, we can pick a particular parameterization of our coordinate space such that the cross ratios can be expressed in the form: 
\begin{equation}\label{eq:DLC-variables}
	\sqrt u = \frac {2 }{\cosh y + \sigma}, \quad \sqrt v = -\frac {\cosh y - \sigma}{\cosh y + \sigma}
\end{equation}
where $\sigma=( r_1^2+r_3^2+ b^2)/(2r_1 r_3) = \cosh \xi + b^2$. Here, $ (y,\xi) $ parametrize our Lorentz boost ($ y $ is essentially rapidity) and dilatation respectively\footnote{Notice that this is a parametrization of our MASG discussed before. See \cite{Raben:2018rbn} for further discussion.}. The cross ratios can be rewritten in terms of $w\equiv \cosh y$, and $\sigma  \equiv \cosh \zeta = \cosh \xi + b^2$. In this frame, causal relation $ x_{23}^2= x_{14}^2<0$ leads to $\sqrt v<0$, which  requires that  $1<  \cosh \zeta< \cosh y$. For s-channel scattering,  the physical region is given by the constraint $ 1<\sigma<w<\infty\ $.

\begin{figure}
	\centering
	\includegraphics[height=3.5cm]{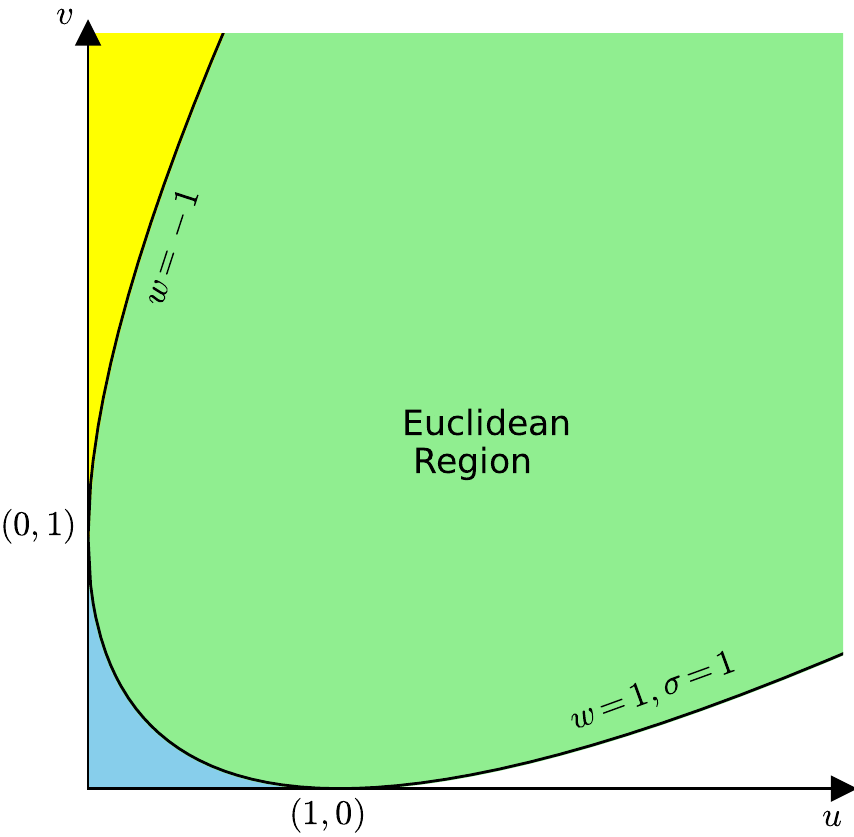}\quad\quad
	\includegraphics[height=3.5cm]{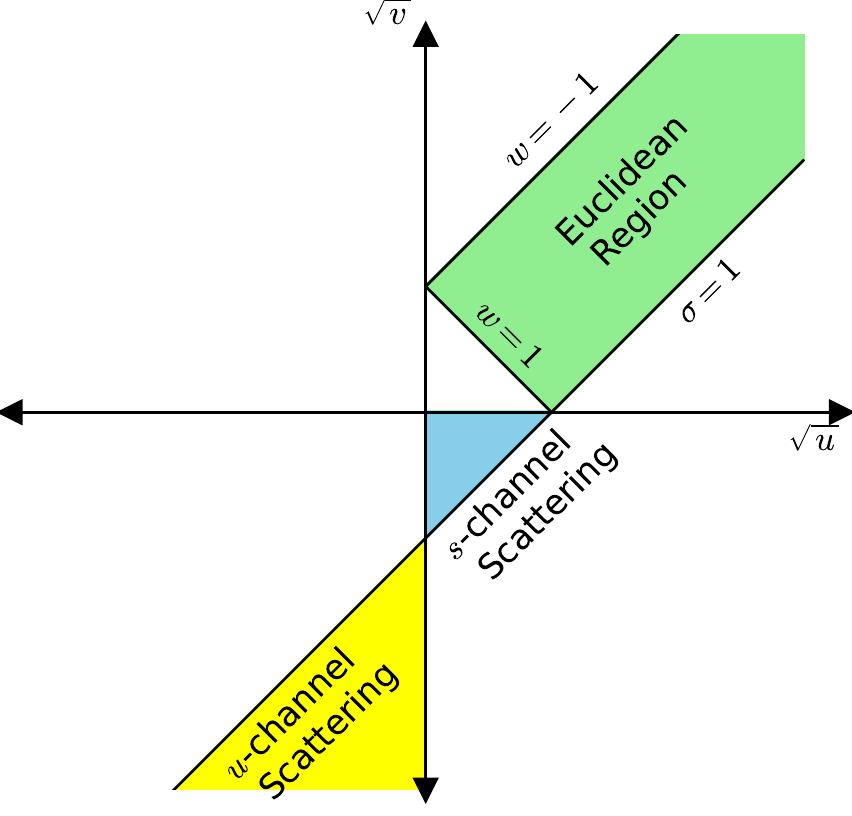}\quad\quad
	\includegraphics[height=3.5cm]{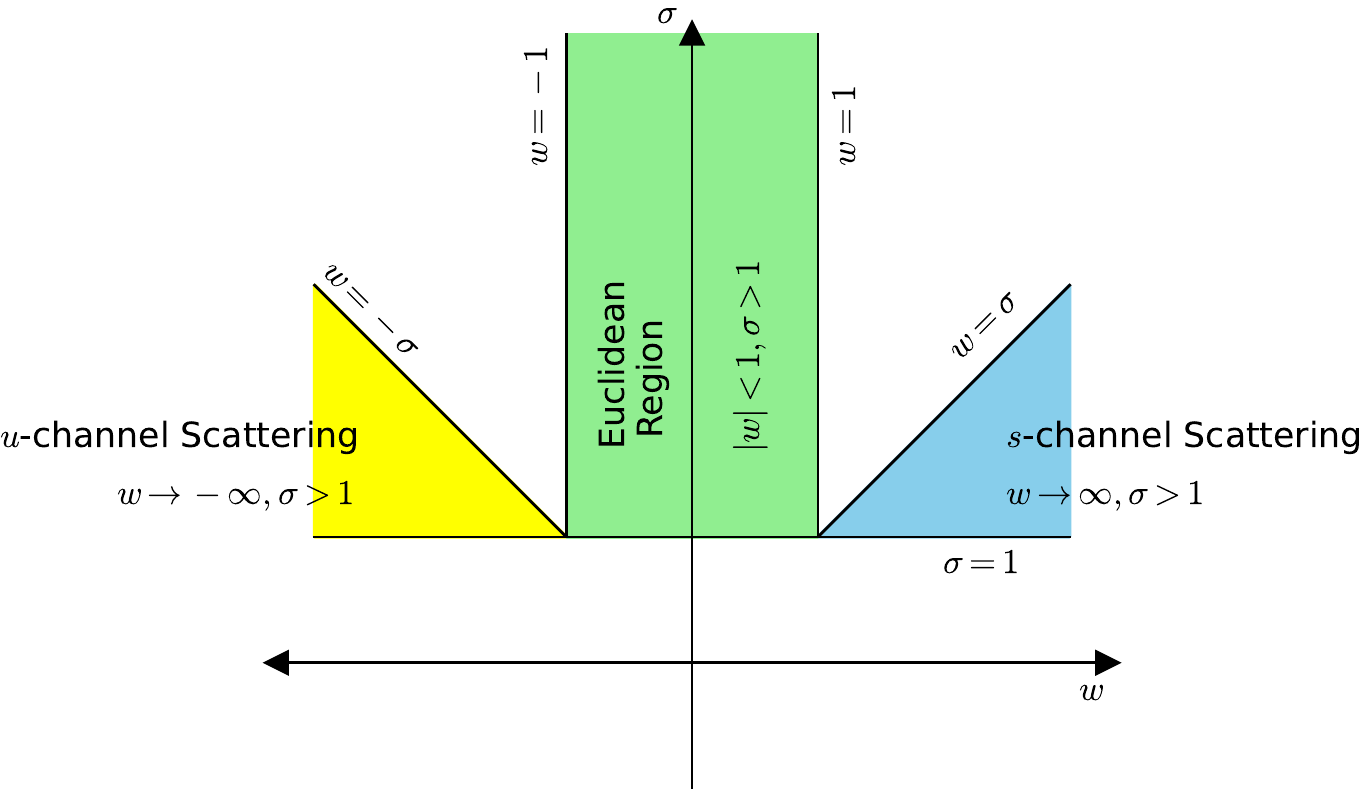}
	\caption{Moving from $ u,v $ plane to $ \sqrt{u},\sqrt{v} $ plane, on to $ w,\sigma $ plane for crossing symmetry. Here, the "Euclidean region" is where the $t$-channel OPE converges as $|w|<1$. This corresponds to $y\rightarrow iy$ in Eq. \ref{eq:DLC-variables}, which is essentially a Wick rotation.}
	\label{fig:crsym}
\end{figure}

\section{Unitary Conformal Blocks}
\label{sec:another}
In order to construct our scattering amplitudes, what we are after is a representation of our symmetry group $ SO(d,2) $. This can be constructed via a direct group theoretic approach: by starting with a $ d+2 $ dimensional embedding space and restricting to a null sub-manifold, we can project down to our $ d $ dimensional spacetime of interest. A representation can then be induced by the MASG $ A $ \cite{knapp2002lie}. In this work, we will focus on an equivalent approach that doesn't quite require the full machinery of group theory by realizing that our partial waves are solutions to the Casimir equation\footnote{A group theoretic treatment involving integrating over the maximal compact subgroup $K$ to obtain these spherical functions  will be presented in \cite{Agarwal:2021}.}.

\subsection{Unitarized Casimir Equation and Conformal Blocks}

In our Minkowski setting, we are interested in the space corresponding to unitary irreducible representation where the dilatation $D$ and Lorentz boost $L_{zt}$ are diagonal, with eigenvalues $\widetilde \Delta$ and $\widetilde \ell$ respectively. For the invariant conformal scalar function of interest to us, the Casimir can be expressed as a differential operator in terms of $(u,v)$\footnote{We have multiplied by a factor of 2 from that defined conventionally, e.g., \cite{Dolan:2011dv}. We have also adopted the sign convention suitable for our Minkowski treatment.}
\begin{equation}\label{eq:Casimir}
	\mathcal{C}=- 2[(1-u-v)\partial_{v}(v\partial_{v})+u\partial_{u}(u\partial_{u}-d)-(1+u-v)(u\partial_{u}+v\partial_{v})(u\partial_{u}+v\partial_{v})].
\end{equation}
It can be shown that  the Casimir acting on the representation space of $SO(d,2)$, labelled by $(\widetilde \ell, \widetilde \Delta)$, can be brought into an explicitly Hermitian form via a Jacobian, taking on the form

\begin{equation}
\widetilde {\cal C}= -\partial_y^2 -\partial_\zeta^2 + V(y,\zeta).
\end{equation}

The Jacobian can be written out explicitly and is chosen to have a simple large $t=e^{y} $  limit, $J \sim t^{-d/2}\sigma^{-(d-2)/2}$.  
With this choice, the potential also takes on a relatively simple form as a function of $(w,\sigma)$ or $(y,\zeta)$ and contains a constant term $V_0$. After resetting the "ground state energy" to be zero, the effective potential $V$ vanishes in the limit $w\rightarrow \infty$, with  $\sigma$ fixed. It follows that the  spectrum is strictly  positive and continuous.  This brings us to one of our main (expected) results:
\begin{equation}
    \widetilde\ell, \widetilde\Delta \rightarrow \text{imaginary}.
\end{equation}
Solutions to  $ (\widetilde {\cal C}-V_0) \Psi(y,\zeta)=\lambda\Psi(y,\zeta)$ are asymptotically plane-wave, with eigenvalues identified with eigenvalues of boost $L_{zt}$ and dilatation $D$, ($k_\xi$ to $k_\zeta$), $\lambda = k_y^2+k_\zeta^2 \equiv  - \widetilde \ell^2  - \widetilde \Delta^2$. It is instructive to compare the eigenvalue for the Casimir, $\lambda+V_0$, to that obtained under the Euclidean treatment, $C(\ell,\Delta) = -[\Delta (\Delta-d) + \ell (\ell +d-2) ]$.
By continuing both $\ell$ and $\Delta$ complex, one finds they can be identified with the corresponding Minkowski values 
\begin{equation}
0<k_y^2= - \widetilde \ell^2 =-(\ell+(d-2)/2)^2, \quad {\rm and}\quad 0< k_\zeta^2 = - \widetilde \Delta^2 =-(\Delta-d/2)^2\ .
\end{equation}
That is, the analytically continued $|\widetilde \ell|$ goes into eigenvalues for the Lorentz boost, $L_{zt}$, and $|\widetilde \Delta|$ into eigenvalues for the dilatation, $D$, in the Minkowski setting, with both $\widetilde \ell$ and $\widetilde \Delta$  purely imaginary. This is indeed our expectation\footnote{These unitarized eigenvalues ($\widetilde \Delta$, $\widetilde \ell$) are related to the root description of our symmetry group $SO(d,2)$  \cite{Agarwal:2021}. Our Jacobian is due to requiring a Unitary Irreducible Representation via the Casimir operator. It is related to the group measure necessary for defining the inner product \cite{10.2307/1969129,knapp2002lie}.}.

A notable feature of this very simple "potential" scattering form of the equation is that it factorizes for the special cases of $ d=2,4 $. There are two ways to write the solution to this equation: as an infinite series of plane wave products or as a product of two hypergeometric functions. We can write the leading behavior as:
\begin{equation}
    G^{(M)}_{(\widetilde\ell,\widetilde \Delta)}(y,\zeta) =J(y,\zeta) {\cal G}_{(\widetilde\ell,\widetilde \Delta)}(y,\zeta) \sim w^{-d/2}\sigma^{-(d-2)/2} ( w^{\widetilde\ell }\sigma^{-\widetilde \Delta })= w^{\ell -1} \sigma^{-\Delta +1}
\end{equation}
where the right hand side is the Regge limit with the expected form \cite{Raben:2018rbn}. 

\subsection{Conformal Dynamics and Meromorphy in $\ell$ and $\Delta$:}
It should be stressed that much of our current study has focussed on the kinematics of unitary representations for $SO(d,2)$. Dynamics of CFT reside in the partial-wave amplitudes $a^{(\alpha)}(\widetilde \ell, \widetilde \Delta)$. In particular, CFT dynamics assume that the singularities consist of poles. Regge asymptotics further requires that these singularities lie on surfaces in the $\widetilde \ell-\widetilde \Delta$ plane, i.e., each surface can be labelled by $\widetilde \Delta = \widetilde \Delta_\alpha(\widetilde\ell)$, $\alpha=1,2,\cdots$, each defining a spectral curve. As discussed in  \cite{Brower:2006ea}, the most important singularity is that interpolating the CFT primary associated with the stress-energy tensor, i.e.,  for $d=4$, $\ell=2$, ($\widetilde \ell= 1$), a pole at $\Delta=4$, ($\widetilde \Delta= 2$). This singularity, at $d=4$, is historically referred to as the Pomeorn. The associated spectral curve can be found in strong coupling via AdS/CFT, leading to a simple parabolic curve.

In order to extract the contribution of the Pomeron at high energies, it is necessary to extend  analytically the representation into the  $\widetilde \ell-\widetilde \Delta$ plane.  This in turn requires one to continue the conformal harmonics analytically. The continuation is analogous to the well-known relation between conventional Legendre functions $
\frac{P_\ell(z)}{\tan \ell \pi} = Q_\ell(z)  -  Q_{-\ell-1}(z)$. This as well as other related issues will be examined in \cite{Agarwal:2021}.

\section{Applications to Diffractive Physics}
The CFT methods described here are most relevant for scattering where the causal structure is important. However, in order to apply these results to QCD studies the conformal symmetry must be broken. This is usually done by working at finite temperature or via a confinement deformation of the dual AdS space: a cutoff is introduced such that the dual space is only asymptotically AdS but has a cutoff in the bulk (at a scale $\sim \Lambda_{\text{QCD}}$). Various thermal, hard wall, and soft wall models for these cutoffs have been developed to great success. 

Our discussion on CFT correlators provides a unified analytic framework that can work directly with both a Minkowski and a Euclidean metric. The causal structure set up in Fig. \ref{fig:crsym} is appropriate for discussing near forward scattering for $t < 0$ and diffractive physics, e.g., examining deep inelastic scattering and inclusive production (e.g. \cite{Nally:2017nsp}). In particular, it will be profitable  to apply this framework to the study of deeply virtual Compton scattering (DVCS) (e.g. \cite{ji1997deeply,Costa:2012fw}). 

\section{Conclusion}
A study of scattering in CFT can provide significant conclusions to be drawn for QCD processes due to its near conformal nature. In a direct group theoretic approach, we expand the $ 2\rightarrow 2 $ scattering amplitudes in a partial wave expansion. These partial waves are unitary irreducible representations of the Lorentzian conformal group $ SO(d,2) $ induced by the maximal abelian subgroup (MASG) $ A=SO(1,1)\times SO(1,1) $. As is typical in a Regge treatment, we are interested in the $t$-channel OPE of the scattering process $ \gamma^* + \gamma^* \rightarrow \gamma^* + \gamma^*$ which in turn lends a natural causal structure to our 4-point functions. With an appropriate gauge choice, we are able to relate the two conformal cross ratios to the 2 parameters of our MASG. By restricting our partial waves to the space of generalized spherical functions and defining an appropriate inner product, we extract the precise leading Regge behavior.



\paragraph{Funding information} P.A. is supported by the National University of Singapore Research Scholarship. This work was supported in part by the U.S. Department of Energy (DOE) under Award No. DE-SC0015845 for R.C.B. T.G.R is funded through an internal research fund at Michigan State University. 



\bibliography{conformald.bib}

\nolinenumbers

\end{document}